\documentclass[preprint2]{aastex}

\slugcomment{}

\shorttitle{Long range outward migration}

\shortauthors{Crida et al.}

\begin{document}

\title{Long range outward migration of giant planets,\\ with
application to Fomalhaut~b}

\author{Aur\'elien Crida}
\affil{Department of Applied Mathematics and Theoretical Physics,
  University of Cambridge,\\ Centre for Mathematical Sciences,
  Wilberforce Road, Cambridge CB3 0WA, \sc United Kingdom}
\email{A.Crida@damtp.cam.ac.uk}

\author{Fr\'ed\'eric Masset}
\affil{Laboratoire AIM-UMR 7158, CEA/CNRS/Universit\'e Paris Diderot,
     IRFU/Service d'Astrophysique,
     CEA/Saclay,
     91191 Gif-sur-Yvette Cedex, \sc France}
\affil{Instituto de Ciencias F\'\i sicas, Universidad Nacional
  Aut\'onoma de M\'exico, Av. Universidad s/n, Cuernavaca, Mor.,
  CP. 62210, \sc Mexico}

\author{Alessandro Morbidelli}
\affil{Laboratoire Cassiop\'ee UMR 6202, Universit\'e de Nice
  Sophia-antipolis / Observatoire de la C\^ote d'Azur / CNRS,\\
  B.P. 4229, 06304 Nice Cedex 4, \sc France}


\begin{abstract}

Recent observations of exoplanets by direct imaging, reveal that giant
planets orbit at a few dozens to more than a hundred of AU from their
central star. The question of the origin of these planets challenges
the standard theories of planet formation. We propose a new way of
obtaining such far planets, by outward migration of a pair of planets
formed in the 10~AU region. Two giant planets in mean motion
resonance in a common gap in the protoplanetary disk migrate outwards,
if the inner one is significantly more massive than the outer
one. Using hydrodynamical simulations, we show that their semi major
axes can increase by almost one order of magnitude. In a flared disk,
the pair of planets should reach an asymptotic radius. This mechanism
could account for the presence of Fomalhaut~b\,; then, a second, more
massive planet, should be orbiting Fomalhaut at about 75~AU.

\end{abstract}


\section{Introduction}

Most of the known exoplanets are gaseous giants, with semi-major-axes
below $2$~AU, or even as close to their parent star as
$0.01$~AU. Inward planetary migration is generally considered as a
necessary phenomenon to explain this distribution, because these
planets should not form there. In the core-accretion model
\citep{Pollack-etal-1996}, giant planets should form beyond the
ice-line (located at about $4$~AU for a solar type star). The general
expectation is that planets form at distances comparable to those
characterizing the orbits of the giant planets of our solar system,
because further out the dynamical (and accretional) timescales are too
long.

However, giant planets have been recently observed by direct imaging
at a few dozens to $120$~AU from their host star
\citep{Kalas-etal-2008,Marois-etal-2008}. It has been proposed that
these large semi-major-axes may be due to scattering with other giant
planets after formation in the $5-20$~AU region
\citep[e.g.][]{Scharf-Menou-2009,Veras-etal-2009}. \citet{Boley-2009},
\citet{Clarke-2009}, and \citet{Rafikov-2009} suggest that planet
formation by gravitational instability \citep[see][for a
review]{Durisen-etal-PPV} can be effective beyond $\sim 50$~AU,
because the cooling time relative to the dynamical time becomes short,
and the Toomre $Q$ parameter may be smaller in the colder, outer parts
of the disk.

In this Letter, we explain how resonant interactions between two
planets in a common gap can lead to outward migration, possibly
explaining the presence of these planets. We outline the mechanism in
Sect.~\ref{sec:review}, and present proof-of-concept simulations in
Sect.~\ref{sec:long}. The effects of disk structure are discussed in
Sect.~\ref{sec:concept}, where we show that a pair of planets can be
driven to an equilibrium radius in a flared disk. Caveats are
discussed in Sect.~\ref{sec:discussion}, and our main conclusions and
applications of this mechanism, paying special attention to the case
of Fomalhaut, are presented in Sect.~\ref{sec:conclu}.

\section{Outward migration in resonance}
\label{sec:review}

Many authors have observed that two planets migrating in a same disk
are often trapped in mean motion resonance
\citep[e.g.][]{Snellgrove-etal-2001,Papaloizou2003,Kley-etal-2004,PierensNelson2008,Crida-etal-2008}.
\citet{MS2001} found that if the outer planet is significantly less
massive than the inner one, and if the two planets open overlapping
gaps, then the migration of the pair can proceed outwards. In this
situation, the inner planet lies close to the inner edge of the common
gap, and far from its outer edge\,; therefore, it feels a positive
torque from the inner disk, and no torque from the outer
disk. Symmetrically, the outer planet mostly feels a negative torque
from the outer disk. If the inner planet is more massive, it feels a
larger torque, in absolute value, than the outer planet, and the total
torque applied to the pair of planets is positive. The pair of planet
therefore moves outwards.

In order for this process to last on the long term, the material lying
outside of the common gap must be funneled towards the inner
disk. Otherwise, it piles up at the outer edge of the gap, which
eventually reverses the torque balance. In addition, the inner disk
must be refilled. In fact, assuming a local damping of the outer
planet's wake, one finds in the simulations an outwards drift rate of
the disk material significantly smaller than that of the
planets. Thus, the gas just outside the gap is necessarily caught up
with the outer separatrix of the outer planet, and thrown inwards.

The whole process may be enhanced by the horseshoe drag due to gas of
the outer disk passing through the common gap --\,like in type~III
migration \citep{MassetPapaloizou2003}. However, the outwards
migration of the pair is observed at all disk mass in numerical
simulations. Since the corotational effects like in type~III migration
become virtually negligible at small disk mass, it is primarily the
wake torque imbalance that drives the outward migration.

Another clue of the importance of the Lindblad torque imbalance on the
process is the sensitivity of the drift rate on the disk's aspect
ratio $H/r$ ($H$ being the scale height of the disk, and $r$ the
distance to the star), as the one-sided Lindblad torque is
proportional to $(H/r)^{-3}$. \citet{MS2001} and
\citet{Morby-Crida-2007} find that the migration rate is a decreasing
function of the aspect ratio of the disk. This effect is further
enhanced by the fact that the tidal truncation of the outer edge of
the common gap by the outer planet is less pronounced in thicker
disks.

\section{Long term simulation}
\label{sec:long}

\subsection{Code presentation}
\label{sec:code}

Accurate numerical simulation of the evolution of giant planets
requires simultaneous computation of planets-disk interactions and
global evolution of the disk. Therefore, we use the 2D1D version
of FARGO \citep{FARGO1,FARGO2}\footnote{\url{http://fargo.in2p3.fr}},
in which the standard 2D polar grid is surrounded by a 1D grid to
compute the disk evolution on all its physical extension
\citep{Crida-etal-2007}.

The 2D grid extends from $0.45$ to $12.61\ L$, where $L$ is the length
unit (taken as $10$~AU). The 1D grid --\,that is, the disk\,--
extends from $0.05$ to $20\ L$. The resolution is $\delta r/r=0.01$,
constant in the two grids. In the 2D grid, $\delta\theta=\delta r/r$.

\subsection{Disk and planets settings}
\label{sub:settings}

The inner regions of protoplanetary disks are poorly constrained,
and disk profiles must be assumed for typical inner disk migration
studies. In contrast, millimeter interferometry
\citep[e.g.][]{Pietu-etal-2007} places strong constraints on disk
structures between $30$ and $200$~AU.

Of particular relevance to scenarios of migration is the temperature
profile, which can be recast as $H/r=c_s/r\Omega$, where $c_s$ is the
sound speed and $\Omega$ the angular velocity. The aspect ratio
follows $H/r=h_0\times(r/L)^\beta$. The flaring index $\beta$ is
typically $0.25$ and ranges from $0.17$ to $0.32$
\citep{Pietu-etal-2007}. The \citeauthor{Pietu-etal-2007} results are
compatible with $h_0=0.045$, for $\mu=2.4$ (mean molecular weight) and
$T=30\,K$ at $100$~AU. Because we use a locally isothermal equation of
state in our simulations, the temperature and aspect ratio are a
function of radius only, independent of time.

For the surface density profile, we assume $\Sigma(r)=
\Sigma_0\times(r/L)^{-p}$. Detection of dust in the continuum
constrains the profile of the dust column density multiplied by its
emissivity (hence its temperature). Assuming a uniform dust to gas
ratio, this yields $1-2\beta + p \approx 2$, thus $p\approx 1.5$,
value adopted in this work \citep[see
e.g.][]{Pietu-etal-2007}. $\Sigma_0$ is much less
constrained. However, unless the disk mass is large enough for
type~III migration to affect the dynamics, our results should be
essentially independent of $\Sigma_0$. More precisely the drift rate
should scale with ${\Sigma_0}^{-1}$\,; this is approximately what is
seen in our numerical simulations with $\Sigma_0$ varying by one order
of magnitude. We therefore adopt values of the surface density that
offer a good trade off between speed and the onset of type~III
migration of the outer planet that would take it away from the
resonance with the inner one.

Calculations are run with a \citet{ShakuraSunyaev1973}
$\alpha$-coefficient between $0.001$ and $0.01$ in order to explore
the importance of the turbulent viscosity $\nu=\alpha c_s H$.

The two planets have masses $M_1=3\times 10^{-3}M_*$ and
$M_2=10^{-3}M_*$, i.e. $2\,M_{\rm Jupiter}$ if $M_*=2M_\odot$.  The
planets are initially on circular orbits at $a_1=L$, $a_2=2L$, and do
not feel the disk potential for the first 100 (Sect.~\ref{sec:result}
and \ref{sec:eos}) or 400 (Sect.~\ref{sec:concept}) orbits of the
inner planet, so the disk can adapt to their presence. Denoting
$r_H=r(M_p/3M_*)^{1/3}$ the Hill radius of a planet, the region inside
$0.6\,r_H$ for a given planet is not taken into account when
calculating the force of the disk on the planet, using a smooth
filter, in agreement with \citet{Crida-etal-2009}.

\subsection{Result}
\label{sec:result}

\begin{figure}
\plotone{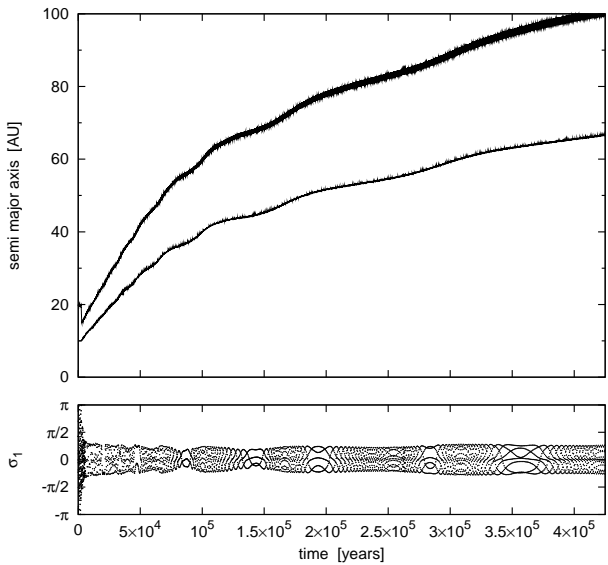}
\caption{Top\,: Semi major axes $a_1$ and $a_2$ as a function of
  time (assuming $M_*=2\,M_\odot$). Bottom\,: Resonant angle
  $\sigma_1=-\lambda_1+2\lambda_2-\omega_1$ as a function of time.}
\label{fig:100}
\end{figure}

Consider the case $\alpha=0.01$, $H/r=0.045(r/L)^{1/4}$, and
$\Sigma_0=1.5\times 10^{-3} M_*L^{-2}$. Assuming $M_*=2\,M_\odot$
(like Fomalhaut) and $L=10$~AU, this is $6.6$ times denser than the
Minimum Mass Solar Nebula \citep{Hayashi1981}\,; the disk parameters
are summarized in Table~\ref{table}. The migration path of the planets
in this disk is plotted in the top panel of Fig.~\ref{fig:100}. After
the release, the outer planet migrates rapidly inwards, and is
captured in 2:1 mean motion resonance. Then, the two planets migrate
smoothly outwards together. The resonant angle
$\sigma_1=-\lambda_1+2\lambda_2-\omega_1$ is displayed in the bottom
panel. The outer planet reaches $10\,L$ ($100$~AU) in $4\times 10^5$
years, and the migration speed decreases as the planets move further
from the star. At distances larger than $100$~AU for the outer
planet, the orbits seem to converge to a stable equilibrium radius.

\begin{table}
\caption{Parameters of the disk used in Sect.~\ref{sec:result}.}
\begin{tabular}{r|cccc}
$r$ & T [K] & $H/r$ & $\Sigma$ [kg.m$^{-2}$] & $Q$ \\
\hline
\hline
$10$ AU & $104$  & $0.045$ & $2666$ & 9.55 \\
$20$ AU & $73.5$ & $0.0535$ & $943$ & 8.0 \\
$50$ AU & $46.5$ & $0.0673$ & $238$ & 6.4 \\
$100$ AU & $32.9$ & $0.08$ & $84.3$ & 5.37
\end{tabular}
\label{table}
\end{table}

Assuming that the migration speed is proportional to the
gas density, the outer planet would reach $100$~AU in $4$ million
years in a disk that is only two thirds the mass of an
MMSN. Because this time is roughly the lifetime of a protoplanetary
disk, a disk that is 10 times lighter than considered above
represents the minimum mass required to migrate the outer planet
beyond 100 AU. Such a disk would be light for a $2\,M_{\odot}$ star,
so we conclude that there is sufficient time for the resonance
mechanism to transport a massive planet to more than $100$~AU.

\section{Extension of concept}
\label{sec:concept}

\begin{figure}
\plotone{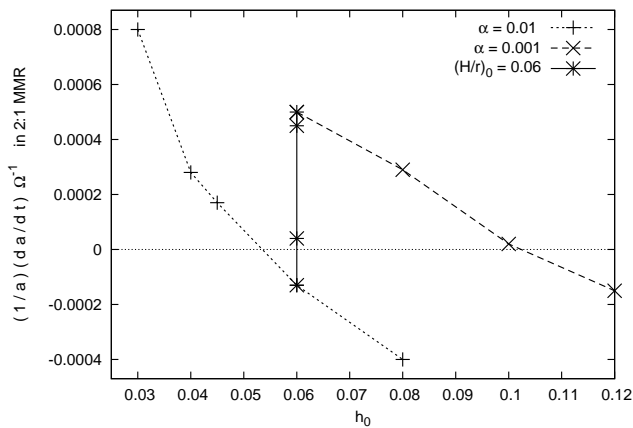}
\caption{Migration speed of a resonant pair of planets, for various
  values of $\alpha$ and $h_0$. $+$ symbols\,:
  $\alpha=10^{-2}$. $\times$ symbols\,: $\alpha=10^{-3}$. Starred
  symbols\,: $h_0=0.06$, and $\alpha=10^{-3}$, $2\times 10^{-3}$,
  $5\times 10^{-3}$, $10^{-2}$ from top to bottom.}
\label{fig:H}
\end{figure}

To understand this process in general, we have performed additional
short-term simulations, using a smaller 2D grid (extending from $0.4$
to $5\ L$), with $\beta=0.25$, $p=1.5$, $\Sigma_0=10^{-3} M_*L^{-2}$
($1778$\ kg.m$^{-2}$ in the case $M_*=2M_\odot$, $L=10$~AU), and with
varying $h_0$ and $\alpha$. In each case, the migration speed of the
pair of planets is measured after the resonance capture. The result is
displayed in Fig.~\ref{fig:H}. The figure is almost filled with
symbols, which means that outward migration is possible for a wide
range of parameters. More precisely, the starred symbols (connected
with the vertical solid line) display the migration rates obtained
with aspect ratio $H/r=0.06 (r/L)^{1/4}$, for various values of
$\alpha$ between $10^{-2}$ (bottom) and $10^{-3}$ (top)\,; ${\rm
d}\ln(a)/{\rm d}t$ is a decreasing function of $\alpha$, and there
exists a critical $\alpha_c$ such that the migration is directed
outward for $\alpha<\alpha_c$ and inward for $\alpha>\alpha_c$.

Reciprocally, for a fixed $\alpha$, ${\rm d}\ln(a)/{\rm d}t$ is a
decreasing function of $H/r$ (see the short-dashed ($\alpha =
10^{-2}$) and long-dashed ($\alpha = 10^{-3}$) curves). In both cases,
there exists a critical aspect ratio $h_c$ such that the migration is
directed outwards for $h_0<h_c$ and inwards for $h_0>h_c$. This
explains partly the slowing down of the migration as the semi major
axes increase in Fig.~\ref{fig:100}, and suggests that in a flared
disk with uniform $\alpha$, the pair of planets should reach the
location where $H/r=h_c$. This is a stable equilibrium point as far as
migration is concerned\,: in the inside, $H/r<h_c$, and migration is
directed outwards\,; further from the star, $H/r>h_c$ and migration is
directed inwards. If $H/r$ increases with $r$, there should be
convergence toward the place where $H/r=h_c$. The evolution of the
density profile also plays a role, but the study of this parameter is
beyond the scope of this letter.

\begin{figure}
\plotone{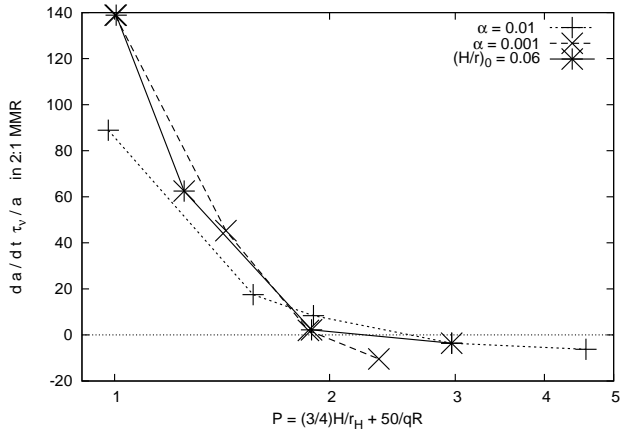}
\caption{Migration speed of a resonant pair of planets, normalized to
  the viscous time, as a function of $P$. Same key as in
  Fig.~\ref{fig:H}.}
\label{fig:P}
\end{figure}

One can expect that the efficiency of the \citet{MS2001} mechanism is
directly related to the size of the gap opened by the outer
planet. To demonstrate this, we plot in Fig.~\ref{fig:P} the same
points and curves as in Fig.~\ref{fig:H}, but the migration speed is
expressed in viscous time $\tau_\nu=r^2/\nu$, and the $x$-axis reports
the parameter $P$, defined as\,:
\begin{equation}
P=\frac{3}{4}\frac{H}{r_H}+\frac{M_*}{M_p}\frac{50\nu}{r^2\Omega}\ .
\label{eq:P}
\end{equation}
This parameter is precisely related to the width and depth of the gap
of the outer planet \citep{Crida-etal-2006}. The strong correlation is
obvious. For a broad range of $H/r$ and $\alpha$, the dependence of
the migration speed on these two parameters can be approximated by a
dependence on the sole parameter $P$. The migration rate is close to
stationary for $2\lesssim P \lesssim 2.5$, whatever the values of
$\alpha$ and $H/r$ that combine into this value of $P$.

\begin{figure}
\plotone{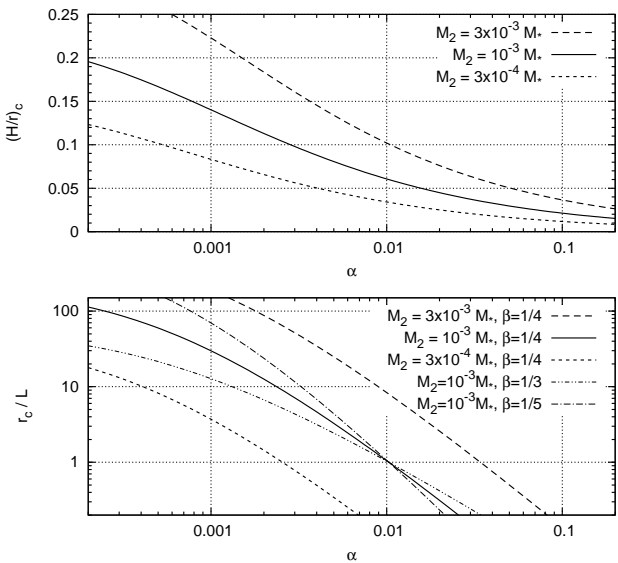}
\caption{Top\,: Solutions of the equation $P=2.5$ for three values of
$M_2/M_*$. Bottom\,: Corresponding $r_c$, for $h_0=0.06$.}
\label{fig:h_c}
\end{figure}

Then, using Eq.~(\ref{eq:P}), $P=2.5$ gives the critical aspect ratio
that halts the migration as a function of $\alpha$, or $\alpha_c$ for
a given $H/r$. The top panel of Fig.~\ref{fig:h_c} shows the value of
the aspect ratio that makes $P=2.5$ for the outer planet, as a
function of $\alpha$, for $2\times 10^{-4}<\alpha<0.2$. The radius at
which the outer planet would stop is then
$$r_c = \left[\frac{(H/r)_c}{h_0}\right]^{1/\beta}L\ .$$
This is displayed in the bottom panel for $h_0=0.06$, so that $L$ can
be understood as the radius in the disk where $H/r=0.06$. The
equilibrium radius $r_c$ depends on $\alpha$, $\beta$, and
$M_2$. Since $\beta$ is generally small, $r_c$ can be very large with
respect to $L$.

Note that for a given value of $P$, the migration speed may depend on
other parameters. Therefore, the critical value for stationary
migration $P\approx 2.5$ holds only in the case considered here, that
is with $M_1/M_2=3$, $\beta=0.25$, $\Sigma_0=10^{-3} M_*L^{-2}$ and
$p=1.5$. A full analysis of all the parameter space is beyond the
scope of this letter.

\section{Discussion}
\label{sec:discussion}

\subsection{Gas accretion on the planets}
\label{sec:accretion}

A shortcoming of the mechanism presented here is that it disregards
the accretion of gas onto the planets. As the pair of planets proceeds
outwards, the gas proceeds inwards through the common gap. The outer
planet should presumably accrete most of the incoming material,
narrowing the mass difference with the inner planet, so that the
torque balance could cancel or reverse at some point. Accretion of the
nebular gas onto a giant planet is a complex process. It is dependent
on the global structure of the flow, on the local properties of the
circum-planetary disk or envelope (poorly described in numerical
simulations), and on the micro-physics of the nebula, such as the
opacity of the grains. Most detailed studies of gas accretion onto
giant planets can't be applied to the present case, because the
conditions are different from the conditions that prevail at $10$~AU,
where these studies apply. This appeals for a thorough investigation
of gas accretion in the conditions of outward migration in the
outskirts of the disk, which is beyond the scope of this work.

\subsection{Self-gravity of the gas and 3D effects}

The gas self-gravity may yield, depending on the equation of state of
the fluid and the mass of the disk, to a vertical compression of the
fluid in the vicinity of the shock
\citep[e.g.][]{Boley-Durisen-2006}. This would increase the Lindblad
torque, more efficiently on the side of the most massive planet, thus
it would actually enhance the mechanism of outward migration. But 3D
effects apply mostly at high order resonances, close to the planet,
which are depleted in our case. In addition, three dimensional
semi-analytic calculations performed by \citet{Tanaka-etal-2002} show
that the angular momentum is essentially exchanged with waves that
have no vertical structure. Therefore, owing to the size of the gap
that our pair of planets carves in the disk, and owing to the
relatively large value of the Toomre parameter (see
Table~\ref{table}), we feel confident that non-self-gravitating, two
dimensional isothermal simulations capture the main features of our
mechanism and provide an acceptable order of magnitude of the outward
drift rate.

\subsection{Equation of state}
\label{sec:eos}

In the above simulations, the equation of state was locally
isothermal. This is realistic at large distances from the star, where
the opacity is low and the cooling time is short with respect to the
dynamical time. However, at $\sim 10$~AU from the star, the cooling
time can be larger than the orbital period. This can affect the aspect
ratio, because of the heating of the disk by the wakes launched by the
planets.

Additional simulations have been run, with the settings of
Sect.~\ref{sec:result}, and computing the full energy equation, with
viscous heating $Q_+$, and no heat diffusion in the disk plane, but
vertical radiative cooling $Q_-=2\sigma_R T^4 / \kappa\Sigma$ (where
$\sigma_R$ is the Stefan-Boltzmann constant). The opacity $\kappa$ is
constant and tuned for the unperturbed disk to be in thermal
equilibrium \citep[see Eq.~(6) of][]{Crida09}. This gives a cooling
time of $111\,\Omega^{-1}$ initially. The aspect ratio increases to
about $0.06$ in the planets region after 100 orbits of the inner
planet. This makes the migration rate after the resonance capture to
decrease to $\sim 10^{-4}$~AU/year instead of $\sim 4\times
10^{-4}$~AU/year.

With a realistic opacity, given by \citet[][Appendix]{BL1994}, $T$ and
$H/r$ are increased for $r \lesssim 10$~AU, and shrink for $r\gtrsim
15$~AU because we don't take heating from the star into account. At
$10$~AU, the cooling time is then $\sim 400\,\Omega^{-1}$. Migration
proceeds outwards at a speed initially similar to the locally
isothermal case (which was expected as $H/r$ is not much changed
around $10-15$~AU), slightly accelerating while reaching the outer
regions where $H/r$ is smaller.

So, it seems that migration can proceed outwards on the long range
also in non locally isothermal disks, even if a self-consistent
radiative disk model still has to be tested.

\section{Applications}
\label{sec:conclu}

We find that a pair of planets formed in the $5-20$~AU region can
reach large semi major axes, if the outer planet is lighter than the
inner one and they orbit in resonance in a common gap in the
protoplanetary disk. The pair of planets tends to reach regions in the
disk where $H/r$ has a critical value (of the order of $0.1-0.15$,
depending on the other parameters)\,; this can be ten times further
from the star than where they formed. After the disk has dissipated,
the evolution of the planets may lead to the disruption of their
original resonant configuration, due to various processes\,: (i) the
eccentricities can rise too much if the planets are too massive\,;
(ii) a third planet could destabilize the system at some
point\,; (iii) the outer planet may enter a debris disk, in which it
migrates independently of the inner planet by planetesimals
scattering. Such a phenomenon may account for the HR\,8799 system,
where the outermost planet is the lightest one, but no resonance has
been identified so far \citep{Marois-etal-2008}.

Otherwise, the planets should remain on orbits in resonance far from
the star. Their eccentricities should therefore be non zero but
moderate\,: during the outward migration, $e$ is excited by the
resonance and simultaneously damped by the disk
\citep[e.g.][]{Crida-etal-2008}. In the simulation presented in
Fig.~\ref{fig:100}, the eccentricities are about $0.01$ and
$0.02-0.03$ for the inner and outer planet respectively. In contrast,
the scattering of giant planets leads to very eccentric orbits (up to
$e=0.8$). The typical final eccentricity of planets formed by
gravitational instability is not known.

The mechanism presented here could account for the case of
Fomalhaut~b. Indeed, this planet has a mass smaller than $1.5\times
10^{-3}M_*$ ($3M_{\rm Jupiter}$)
\citep{Kalas-etal-2008,Chiang-etal-2009}, while \citet{Quillen-2006}
predicted a planet smaller than Saturn. Thus, a more massive, inner
planet could exist. In addition, \citet{Kalas-etal-2008} notice that
Fomalhaut~b is not apsidally aligned with the dust belt, which
suggests the presence of additional perturbers in the
system. Observations in the $M$-band by \citet{Kenworthy-etal-2009}
rule out the presence of a planet more massive than $2\ M_{\rm
Jupiter}$ in the range $8-40$~AU from Fomalhaut. However, if our
mechanism occurred in the Fomalhaut system, then the second planet
should be orbiting at about 75 AU from the star, with a mass of the
order of $1$ to $10$ Jupiter masses.

\acknowledgments

\small

Simulations have been performed on the {\tt hyades} cluster of the
D.A.M.T.P., on the on the {\tt hpc-bw} cluster of the {\it
Rechenzentrum} of the University of T\"ubingen, and on a 92 core
cluster funded by the program ``{\it Origine des Plan\`etes et de la
Vie}'' of the French {\it Institut National des Sciences de l'Univers}.
A. Crida acknowledges STFC and J. Papaloizou.
We acknowledge enlightening discussions with V. Pi\'etu.
We also thank the anonymous referee for his/her insightful remarks,
which improved the paper.

\end{document}